# The case of "Less is more"

## Modelling risk-preference with Expected Downside Risk

**MIHÁLY ORMOS**
Department of Finance
Budapest University of Technology and Economics
Magyar tudosok krt 2., Budapest, Hungary
ormos@finance.bme.hu

**DUSÁN TIMOTITY**
Department of Finance
Budapest University of Technology and Economics
Magyar tudosok krt 2., Budapest, Hungary
timotity@finance.bme.hu

### Abstract

This paper discusses an alternative explanation for the empirical findings contradicting the positive relationship between risk (variance) and reward (expected return). We show that these contradicting results might be due to the false definition of risk-perception, which we correct by introducing Expected Downside Risk (EDR). The EDR parameter, similar to the Expected Shortfall or Conditional Value-at-Risk, measures the tail risk, however, fits and better explains the utility perception of investors. Our results indicate that when using the EDR as risk measure, both the positive and negative relationship between expected return and risk can be derived under standard conditions (e.g. expected utility theory and positive risk-aversion). Therefore, no alternative psychological explanation or additional boundary condition on utility theory is required to explain the phenomenon. Furthermore, we show empirically that it is a more precise linear predictor of expected return than volatility, both for individual assets and portfolios.

**Keywords**: asset pricing; variance; conditional value at risk; expected downside risk; utility theory; behavioral finance
**JEL classification**: G02; G12 G17; C53; C62

**Acknowledgements:** We would like to gratefully acknowledge the valuable comments and suggestions of two anonymous referees that contribute to a substantially improved paper. Mihály Ormos acknowledges that this study was supported by the János Bolyai Research Scholarship of the Hungarian Academy of Sciences

# 1. Introduction

According to Modern Portfolio Theory (MPT) (Markowitz 1959), if returns follow elliptical distributions, the utility of stochastic investment opportunities can be described as the following, by assuming either wealth-dependent (constant relative, CRRA) or wealth-independent (constant absolute, CARA) risk-aversion under Expected Utility Theory (EUT) (Gossen 1854). In either case, the Taylor approximation of expected utility yields the following relationship between expected return and variance:

$$U(F) \cong E(F) - 0.5a\sigma^2 \qquad (1)$$

where *U(F)*, *E(F)* and *σ* stand for the utility, expected value and standard deviation of a the possible realizations of a one-period investment, while *a* is an Arrow-Pratt measure of absolute risk-aversion. By assuming unit wealth, this well-known equation in asset pricing (e.g. Capital Asset Pricing Model by Lintner 1965; Mossin 1966; Sharpe 1964) suggests a positive relationship between expected nominal changes (or if unit wealth is assumed, expected returns) and volatility under standard circumstances (i.e. positive risk-aversion coefficient) Therefore, regardless of the utility function used (CARA or CRRA), equation (1) holds.

However, contradicting empirical results (Brooks et al. 2014; Kahneman and Tversky 1979, Tversky and Kahneman 1992; Linville and Fischer 1991; Post and Levy 2005) indicate a negative relationship between the two parameters leading to the emergence of an alternative utility theory: the prospect theory of Kahneman and Tversky (Kahneman and Tversky 1979). The authors find that in certain cases (e.g. decision on losses) if investors have two options yielding the same expected return they tend to choose the option involving higher risk. They propose a convex, fluctuation dependent utility function for losses as a solution to the problem.

In contrast, we argue that this behavior deviating from that suggested by the EUT is not necessarily due to a flaw in the theory itself but the measurement of the risk perception. Although numerous novel risk measures have emerged since the MPT, such as Value-at-Risk (VaR) (Campbell et al. 2001; Jorion 2007) or Conditional Value-at-Risk (CVaR) (Rockafellar and Uryasev 2000; Acerbi and Tasche 2002) or entropy (Ormos and Zibriczky 2014), none of them had the initial purpose of contributing to equilibrium modelling by describing risk-preference in a more precise way. Therefore, we introduce Expected Downside Risk (EDR) (Ormos and Timotity, 2016a, 2016b), based on CVaR, with all its advantages but without its disadvantage of using a pre-defined probability level. Our proposed measure is effectively the expectation of returns below the expected return, or in other words, the expected bad outcome. Hence, in contrast to standard risk-measures (e.g. volatility or VaR), higher EDR means lower risk, and requires lower expected return as compensation for risk-aversion.

Subsequent to defining our proposed risk measure and its theoretical application in asset pricing, we provide an empirical analysis on its explanatory power of the average return of individual assets as well as well diversified portfolios, test whether the theoretical findings hold in reality as well and make comparison with its peers such as volatility, variance and 5% Conditional Value-at-Risk. As the main contribution of this paper, we show that, without leaving the standard EUT framework, both the theoretically optimal portfolio and the empirically best fitting linear model can reflect both negative and positive risk-return relationship in our proposed model.

The rest of the paper is organized as follows: in Section 2 we introduce and discuss our proposed method for measuring risk perception, in Section 3 the empirical investigation of our proposed explanation is presented and lastly in Section 4 we provide a brief conclusion.

## 2. The Model

We argue that the flaw in the conclusion of contradicting experimental and empirical results on the relationship between risk and return comes from the fact that the tests try to explain the results in a volatility or variance based setting. This setting, as suggested by equation (1), indeed yields the result that investors having dominantly positive risk-aversion coefficient (Barsky et al. 1997; Hanna and Lindamood 2004) should choose an investment option involving a lower risk for a given level of expected return.

In contrast, in line with prospect theory, we argue that investors do not focus on (in practical terms, they do not perceive) the volatility or variance of their investment but the expected loss of it. However, we do not assume that investors should behave according to Prospect Theory, that is, they do not necessarily follow loss aversion instead of risk aversion. We argue that keeping the EUT setting with constant risk-aversion while changing the applied risk measure, one may explain the contradicting results of risk- and loss-aversion.

This change of risk measure also allows for including other properties of the return distribution (e.g. skewness) that have been documented to play an important role in explaining the risk premium (Astebro et al. 2014; Post et al. 2008). There are some important risk measures that cope with asymmetric distributions, and still can be used as coherent measures (Artzner et al. 1999), and are supported by general equilibrium (Csoka et al., 2007), such as the Conditional Value-at-Risk, but are defined with an ad-hoc probability level that questions their robustness.

Our proposed measure is the expected loss of an investment labelled as the Expected Downside Risk (EDR), which is defined in the following way: similar to the Conditional Value-at-Risk, *EDR* measures the expected tail risk; however, it does not apply an ad-hoc probability level and considers returns below the expected return as a loss, that is

$$EDR(x) = p(r(x) \leq E(r(x)))^{-1} \int_{r_x(y) \leq E(r(x))} r_x(y) p(y) dy. \qquad (2)$$

where $r(x)$ and $E(r(x))$ sign the return and its expectation of a given portfolio x and $r_x(y)$ and $p(y)$ stand for the outcomes of the portfolio *x* and their respective probabilities. In other words, *EDR*, as its name suggests, measures the expected loss (risk) of investors given that their reference point is the expected return as suggested by Easterlin (1974) or Kőszegi and Rabin (2006).

By looking at the aforementioned definition, one may find EDR very similar to Expected Shortfall (ES) or Conditional Value-at-Risk (CVaR). In fact, depending on skewness of the return distribution, EDR is equal to the CVaR calculated with the quantile of the median with an opposite sign. For example, for symmetric distributions, where the median and the mean are located at the same point, EDR is always defined as the opposite of the 50% CVaR.

This change of risk measure is a plausible modification of standard asset pricing models assuming EUT for two reasons. On the one hand, due to bounded rationality (Simon 1982), measures being easier to interpret are the main factors driving the focus of investors (Gigerenzer and Selten 2002); in particular, this is an important reason why fixed monetary payments and assigned probabilities are applied in experimental and laboratory tests instead of complicated mathematical formulas (Tversky and Kahneman 1992), such as a variance-expected return choice set. Moreover, the precise prediction of such measures for the future requires even more complex techniques (Andor and Bohák, 2016; Ormos and Timotity, 2016c). Therefore, the expected amount of money an investor can lose might be of greater relevance than the expectation of the squared deviations from the mean. This idea is in line with recent studies finding that separately applying losses and gains in asset pricing models yield higher goodness-of-fit (Tsai et al., 2014; Cheng et al., 2014). On the other hand, the variance based approach yields the optimal portfolio choice only if each compared investment has the exact standardized distribution (i.e. excluding the variance and mean, they are identical). This problem has already been solved by Value-at-Risk and Expected Shortfall due to their non-parametric approach, however, both have a pre-defined probability-level that cannot be verified fundamentally (i.e. the reason for the choice of a given probability level does not have an economic explanation). Nonetheless, in our empirical results section we compare *EDR* against these risk measures as well using the most popular probability levels.

Assuming that, apart from the expected return and variance, the distribution type of the return of a given asset does not change over time (which, in reality, is a fairly valid assumption (Singleton and Wingender 1986; Sun and Yan 2003)) its *EDR* can be described as a linear function of its expected return and volatility of past return. In order to illustrate this relationship we provide an example with normal distribution as approximation, which allows for tighter conditions; however real distributions of returns can be calculated in the same way by changing the coefficient only. Below, we provide our first theorem and its proof that derives the risk-return relationship in the asset pricing model based on the EUT.

**Theorem 1:** Indifference curves are quadratic functions of the EDR and the expected return of a distribution.

**Proof:** We can define EDR as the function of expected return and standard deviation:[1]

$$EDR(x) = E(r(x)) - 0.8\sigma \qquad (3)$$

According to this equation, we can substitute the volatility ($\sigma$) with EDR in equation (1). Therefore, the approximating function can be implemented in the EDR-E(r) setting:

$$U = E(r(x)) - \frac{0.5}{0.8^2}a[E(r(x)) - EDR(x)]^2. \qquad (4) \square$$

---

[1] in the case of normal distribution EDR=-CVaR$_{0.5}$= $\int^{E(r)} r \cdot \left(\frac{1}{\sigma\sqrt{2\pi}} e^{\frac{[r-E(r)]^2}{2\sigma^2}}\right) dr$ that is equal to 0.8 assuming standard normal distribution, and therefore, adding a constant (the expected return) and a multiplication by the standard deviation yield equation (3)

In Figure 1 we provide a numerical simulation of the indifference curve for an average constant absolute risk-aversion coefficient (Barsky et al. 1997) of *a*=4, where points signed by +, - and *x* represent utility levels of U=1, 2 and 3 respectively. Here, we underline that this generally accepted level of absolute risk-aversion is approximately equal to the relative risk-aversion measure of investors behaving in a similar pattern; therefore, again, one can use both utility functions in the equation above. Although these examples use normal distribution with the 0.8 volatility coefficient seen in equation (3), we show later that in general the coefficient is actually not far from this value.

**Please insert Figure 1 here**

Furthermore, the generalized form of equation (4) yields

$$0 = -c\left[\left(E(r(x)) - \frac{2cEDR(x)+1}{2c}\right)^2\right] + EDR(r(x)) + \frac{1}{4c} - U, \quad (5)$$

where *U* stands for the utility and *c* is a constant for each investor defined as

$$c = \frac{0.5}{0.8^2}a. \quad (6)$$

Equation (5) clearly indicates the quadratic indifference curve, however, it also suggests a more intuitive result: since the first term is always non-positive (assuming a positive risk-aversion coefficient) the following equation must hold in order to have a real solution, that is

$$EDR(x) + \frac{1}{4c} - U \geq 0. \tag{7}$$

For further analysis of our asset pricing model, we define the slope of the indifference curve in the *EDR-E(r)* setting. In the following theorem, we highlight the main theoretical contribution of our paper.

**Theorem 2:** In the EDR-E(r) system, both a negative and a positive relationship between risk and return can be derived under standard, EUT conditions.

**Proof:** In the followings, we use the general, parametric approach, hence, the deduction is valid for any distribution. In this generalization, instead of the 0.8 level, we assign to *v* a distribution-dependent coefficient of the volatility, that is $v = \frac{E(r(x)) - EDR(x)}{\sigma}$. We know that the total derivative of the indifference curve should be zero; therefore, the slope can be calculated as in equation (8).

$$\frac{dU}{dEDR(x)} = \frac{a}{v^2}\big(E(r(x)) - EDR(x)\big) + \frac{dE(r(x))}{dEDR(x)}\left(1 - \frac{a}{v^2}\big(E(r(x)) - EDR(x)\big)\right) = 0. \tag{8}$$

From equation (8), the sensitivity of the expected return for Expected Downside Risk can be expressed as

$$\frac{dE(r(x))}{dEDR(x)} = -\frac{\frac{a}{v^2}\big(E(r(x)) - EDR(x)\big)}{1 - \frac{a}{v^2}\big(E(r(x)) - EDR(x)\big)} = 1 - \frac{1}{1 - \frac{a}{v^2}\big(E(r(x)) - EDR(x)\big)}. \tag{9}$$

This relationship implies that if $EDR(x) < E(r(x)) - \frac{v^2}{a}$, the slope of the indifference curve is positive and greater than one

$$\frac{dE(r(x))}{dEDR(x)} > 1, \tag{10}$$

if $EDR(x) = E(r(x)) - \frac{v^2}{a}$, the slope is positive or negative infinity, and if $EDR(x) > E(r(x)) - \frac{v^2}{a}$, the slope can be both negative and positive, but less than one

$$\frac{dE(r(x))}{dEDR(x)} < 1. \tag{11}$$

Furthermore, the expectation of returns not greater than the expected return cannot be higher than the expected return itself; hence, including the constraint of $EDR(x) \leq E(r(x))$,

$$E(r(x)) - \frac{v^2}{a} < EDR(x) \leq E(r(x)) \tag{12}$$

yields the "usual" positive relationship between risk and expected return. However, in equation (10) we have shown that in the case of small EDRs the relationship changes and a higher risk will be rewarded with lower expected return. □

The aforementioned relationship is presented in Figure 2, where the curve, the dashed line and the dotted line stand for the indifference curve, the risk-free portfolios and the $EDR(x) = E(r(x)) - \frac{v^2}{a}$ constraint respectively. Here we use *a*=0.5 and *v*=0.8 parameters.

**Please insert Figure 2 here**

Since the appearance of the CAPM, asset pricing models also implement the effects of leverage opportunities at the risk-free rate. Therefore, in the followings, we also show that the aforementioned, changing relationship between risk and reward shows up in the leveraged portfolio optimization as well.

**Theorem 3:** If leverage is included in the EDR-E(r) setting, portfolio optimization could yield both negative and positive slopes for the "capital market line" collecting the efficient portfolios.

**Proof:** In order to analyze the case including leverage opportunity, first we have to define whether it provides an optimal solution as well. Here, we assume a unique risk-free asset that provides risk-free return $r_f$. In line with the standard volatility – expected return-based Markowitz model or the Beta-based CAPM, the weight of risk-free asset in the portfolio affects linearly both the EDR and the E(r). Therefore, the expected return of the leveraged portfolio can be defined as a linear function of EDR, where the ratio between risk and return or the price of risk ($k$ in equation (13)) is constant, which is similar to the security market line of the CAPM. This finding further implies that, if an optimal solution exists, it is equal to the point of tangent of the leveraged portfolios of

$$E(r(x)) = r_f + kEDR(x), \tag{13}$$

and the highest indifference curve

$$U = E(r(x)) - \frac{0.5}{v^2} a[E(r(x)) - EDR(x)]^2. \tag{14}$$

Based on investors' goal of maximizing their utility we get a simple linear constraint optimization problem from equation (13) and (14), which is

$$\max_{E(x), EDR(x)} \left\{ E(r(x)) - \frac{0.5}{v^2} a[E(r(x)) - EDR(x)]^2 \right\} \text{ s.t. } E(r(x)) = r_f + kEDR(x). \tag{15}$$

Using the Lagrangian and its derivatives we get

$$L = E(r(x)) - \frac{0.5}{v^2} a[E(r(x)) - EDR(x)]^2 + \lambda \left( E(r(x)) - r_f - kEDR(x) \right), \tag{16}$$

$$\frac{dL}{dE(r(x))} = 1 - \frac{a}{v^2} [E(r(x)) - EDR(x)] + \lambda = 0, \tag{17}$$

$$\frac{dL}{dEDR(x)} = \frac{a}{v^2} [E(r(x)) - EDR(x)] - \lambda k = 0, \tag{18}$$

$$\frac{dL}{d\lambda} = E(r(x)) - r_f - kEDR(x) = 0. \tag{19}$$

According to eq. (17)-(18) the optimal solution for the constrained optimization problem is

$$EDR(x)_{opt} = \frac{kv^2}{a(k-1)^2} - \frac{r_f}{k-1}, \qquad (20)$$

$$E(r(x))_{opt} = \frac{kv^2}{a(k-1)} + \frac{kv^2}{a(k-1)^2} - \frac{r_f}{k-1}. \qquad (21)$$

We further have the condition of $E(r(x))_{opt} \geq EDR(x)_{opt}$, therefore, the solution is valid if and only if

$$\frac{kv^2}{a(k-1)} \geq 0 \qquad (22)$$

$$k \in (1, \infty) \cup (-\infty, 0]. \qquad (23)$$

In order to define the optimal portfolio that is used in leveraging we determine the slope coefficient in the following way. Substituting back into eq. (14) we get

$$U = \frac{kv^2}{a(k-1)} + \frac{kv^2}{a(k-1)^2} - \frac{r_f}{k-1} - \frac{0.5}{v^2} a \left[\frac{kv^2}{a(k-1)}\right]^2 = \frac{v^2k - 0.5v^2k^2}{a(k-1)^2} + \frac{v^2k - ar_f}{a(k-1)}. \qquad (24)$$

Then the derivative of (22) is

$$\frac{dU}{dk} = \frac{ar_f(k-1)-v^2 k}{a(k-1)^3} = \frac{(ar_f-v^2)k-ar_f}{a(k-1)^3} = 0, \tag{25}$$

we find that

$$\frac{d^2 U}{dk^2} = \frac{2ar_f(1-k)+v^2(2k+1)}{a(k-1)^4} = \frac{-2(ar_f(k-1)-v^2 k)+v^2}{a(k-1)^4}, \tag{26}$$

which is always positive under the first order condition (25), therefore

$$\frac{d^2 U}{dk_0^2} = \frac{v^2}{a(k_0-1)^4} > 0 \tag{27}$$

if $\frac{dU}{dk_0} = 0$ where $k_0 = \frac{ar_f}{ar_f-v^2}$ stands for the extremum of the slope coefficient. These together yield that the utility function does not have a local maximum but a minimum (in line with Figure 1 and Figure 2). Furthermore, as the derivative of the leveraged utility function with respect to the slope coefficient depends on both the slope and the given risk-aversion, risk-free return and volatility combination, we define the following cases:

If $ar_f > v^2$

$$\frac{dU}{dk} \begin{cases} > 0 & if \quad k < 0 \\ < 0 & if \quad 1 < k < \frac{ar_f}{ar_f-v^2} \\ > 0 & if \quad k > \frac{ar_f}{ar_f-v^2}, \end{cases} \tag{26}$$

and if $ar_f < v^2$

$$\frac{dU}{dk} \begin{cases} < 0 & if \quad k < \frac{ar_f}{ar_f - v^2} \\ > 0 & if \quad \frac{ar_f}{ar_f - v^2} < k < 0 \\ < 0 & if \quad k > 1. \end{cases} \quad (27)$$

These are the main theoretical findings of the proposed model. They indicate that depending on the exogenous parameters of $a, r_f, v^2$ (i.e. the risk-aversion coefficient, the risk-free rate and the coefficient of volatility in the *EDR* regression) the leveraged optimization could lead to both negative and positive slope choice. Investors maximize utility either by choosing the leverage slope closest to the *EDR=E(r)* line or by they picking the leveraged portfolio line the furthest from the 45 degree line.

The theoretical maximum of utility would be reached at the leveraged line with slope closest to unity since the limit from the right (getting closer to the *EDR-E(r)* line on the positive side)

$$\lim_{k \to 1+} U = \lim_{k \to 1+} \left[ \frac{v^2 k - 0.5 v^2 k^2}{a(k-1)^2} + \frac{v^2 k - ar_f}{a(k-1)} \right] = \infty \quad (28)$$

is infinite, while the leverage portfolio line the furthest yield only a finite utility since

$$\lim_{k \to 0} U = \lim_{k \to 0} \left[ \frac{v^2 k - 0.5 v^2 k^2}{a(k-1)^2} + \frac{v^2 k - ar_f}{a(k-1)} \right] = \frac{v^2}{2a}. \quad (29)$$

Therefore, investors would theoretically prefer the leveraged line with slope of unity over the one with the slope of zero, however, in the real world these portfolios are not always attainable for investors. In these cases, the leveraged portfolio line with the highest negative slope coefficient might be the optimal choice.                                                          □

Figure 3 summarizes the aforementioned equations where line 1 and line 2 stand for the $EDR \leq E(r)$ constraint and the $k = \frac{ar_f}{ar_f - v^2}$ condition (i.e. the slope where utility is minimal) respectively.

**Please insert Figure 3 here**

## 3. Empirical results

### 3.1 Data

In our empirical investigations we test the realization of the *EDR-E(r)* equilibrium for equity portfolios, moreover, we analyze how the equilibrium is parameterized; first, for unleveraged portfolios, and second, for leveraged ones. We apply daily and annual data and statistics to investigate unlevered and levered portfolios. The data used for these portfolio calculations consist of daily returns from July 31, 1993 to July 31, 2014 of the 340 constituents of S&P 500 index that have been listed both at the beginning and at the end of the period. Thus the dataset we use is not free of survivorship bias. In order to model leveraged portfolios we apply the mean annualized 3-month T-bill log return for the same 21 years. This dataset is used in sections 3.2 and 3.4.

In section 3.3, where no individual asset is considered, we apply longer historical time series: we use the annual returns of the S&P 500 index and the 3-month Treasury bills between 1928 and 2013.

### *3.2 Optimization for unleveraged and leveraged portfolios*

In the case of unleveraged portfolios we simulate the performance of 10,000 randomly weighted portfolios consisting of the 340 different stocks. The *EDR* calculation is based on either daily or yearly returns. The plot of these simulated portfolios using yearly statistics is presented in Figure 4. One can see that the concave efficient frontier representing portfolios with the highest *EDR/E(r)* ratio always has an optimum at the point of the tangent with the convex utility functions (as shown in Figure 1).

**Please insert Figure 4 here**

Our utility maximization results based on the simulated EDR-E(r) pairs indicate that for *a*=0.5 (extremely low) and *a*=4 (average) risk-aversion the {*EDR,E(r)*}={0.15, 0.27} portfolio is optimal on the simulated set of possible portfolios; however, for *a*=10 (extremely high) the {0.15, 0.24} is optimal. Here, we find evidence of the surprising utility preference we have derived in the previous section: for a given level of risk (measured by the *EDR*), a portfolio with a lower expected return may provide higher utility; in particular, the utilities generated by equation (14) are shown in Table 1. The results show two important patterns: on the one hand, for fixed expected return the increasing risk-aversion, as expected, decreases the utility of the stochastic payoff; on the other hand, for fixed risk-aversion, one may clearly see that for *a*=0.5 and *a*=4 a decrease in the expected return yields a loss of utility, which is in line with standard asset pricing theories, however, this relationship is the opposite for *a*=10.

**Please insert Table 1 here**

This latter finding is due to the following mechanism: since EDR is fixed here, a decrease in the expected return comes with a proportional drop in return volatility; initially, at low risk-aversion levels (*a*=0.5, *a*=4), the drop in the expected return hurts more than the excess utility provided by the decreasing volatility; however, at very high risk-aversion levels, the latter becomes larger in magnitude and can counterbalance the loss of expected return. Nevertheless, using the EDR as risk-measure, one can see both increasing (*a*=0.5, *a*=4) and decreasing utility (*a*=10) as a function of the expected return.

We also run the simulation using daily statistics in Figure 5. Using a similar dataset (as above) at a daily level, we find that the optimal portfolios are much further away from each other than in the annual analysis. This is mainly due to the fact that the point of tangent between the efficient frontier and the indifference curve is on the decreasing part of the latter.

**Please insert Figure 5 here**

In case of leveraged portfolios we first apply the slope optimization described in Section 2. In order to illustrate this optimization let us include a risk-free rate in the aforementioned portfolio simulation. Using the mean annualized 3-month T-bill log return for the same 21 years we get a risk-free rate ($r_f$) of 2.73%. In addition, applying the OLS estimation of

$$E(r(x)) - EDR(x) = v\sigma \qquad (28)$$

to the yearly portfolio simulation results yields a fitted volatility coefficient of *v*=0.78. The difference from normality (i.e. the theoretical volatility coefficient of 0.8) is well-reflected here. For comparison purposes we mention that we measure daily, and monthly volatility coefficients to be 0.68 and 0.76 respectively, which are in line with the standard findings that deviation from normality (e.g. fat tails) decreases as the investment horizon increases.

These parameters further lead to the definition of $k_0$ and yield the sign of the EDR-E(r) relationship in the following way: according to the simulated portfolios, the leveraged portfolio lines have the slope of $\frac{dE(r(x))}{dEDR(x)} = 1.64$ and $\frac{dE(r(x))}{dEDR(x)} = 0$ at the boundaries (the closest to and the furthest from the EDR=E(r) line). Utility at the boundaries is obtained by substituting back into eq. (24). Analysis of the utility difference of $U_{k=1.64} - U_{k=0}$ yields that for $\in (0,28.55)$ $U_{k=1.64} > U_{k=0}$, hence, a positive relationship between *EDR* and *E(r)* is preferred. Figure 6 shows the leveraged portfolio optimization conditional on yearly parameters, where the red "+" sign stands for the risk-free asset while the horizontal and $\frac{dE(r(x))}{dEDR(x)} = 1.64$ sloped lines represent the optimal leverage line given $a \geq$ 28.55 and $a < 28.55$ respectively. According to this optimization investors having $a <$ 28.55 (the majority according to Barsky et al. (1997) and Hanna and Lindamood (2004)) choose from the leveraged portfolios on the *EDR=r_f+1.64E(r)* line, while those having $a \geq 28.55$ invest into portfolios providing *EDR=r_f*.

The daily analysis yields somewhat different results. Here, we measure an average $r_f = 0.011\%$, *v*=0.68 and the boundary slopes range from 0 to -0.08. This case implies that for $\in (0,155.69)$ $U_{k=-0.08} > U_{k=0}$. This simulation is shown in Figure 7. It means that practically every investor picks portfolios from the *EDR=r$_f$-0.08E(r)* line, and therefore, we see a negative relationship between *EDR* and *E(r)* at the daily frequency. This negative relationship between risk and reward in the short term is well in line with recent literature, which confirms the existence of investors increasing portfolio risk immediately subsequent to negative asset price shocks (Ormos and Timotity, 2016d; Ormos and Timotity, 2016e).

**Please insert Figure 7 here**

### *3.3 Ambiguous risk-preferences*

Here we illustrate the aforementioned optimization with the following simple portfolio choice problem. We measure the following parameters of a well-diversified portfolio (we assume the S&P500 index behaving as the market portfolio): the average log return $E(r_m) = 9.12\%$, the annual return volatility $\sigma_M = 19.5\%$, $EDR_m = -13.07\%$ and $r_f = 3.47\%$. The *v* parameter introduced in equation (8) hence takes on $v = 1.14$. Based on Barsky et al. (1997) and Hanna and Lindamood (2004) we assume an average investor having a risk-aversion coefficient of $a = 4$.

Now let us look at the boundary portfolio providing minimal *EDR* where equation (9) yields an infinite indifference curve slope. We know that equation (3) and its modification with the volatility coefficient still holds, therefore

$$EDR_P = E(r)_P - \frac{v^2}{a} = E(r)_P - v\sigma_P. \tag{29}$$

Then, equation (30) follows as

$$\sigma_P = \frac{v}{a}. \tag{30}$$

We further know that leveraged positions combined with the risk-free interest rate yield the following equation

$$E(r)_P = r_f + \frac{E(r_m) - r_f}{\sigma_M} \sigma_P. \tag{31}$$

The solution of the equation system yields $E(r)_P = 11.72\%$, $\sigma_P = 0.2846$, $EDR_P = -20.67\%$ and $U_P = -0.0447$. Now let us measure the indifference curve of the same investor by testing the *EDR-E(r)* pairs with "decreased" risk. Here we mean decreased risk by increasing EDR for a given utility level, that is, let us consider a portfolio X where $EDR_X = -10\%$. By the definition of the indifference curve $U_P = U_X$, which gives two possible solutions of $E(r)_X = \begin{Bmatrix} -3.89\% \\ 48.68\% \end{Bmatrix}$ with $\sigma_X = \begin{Bmatrix} 0.0472 \\ 0.4529 \end{Bmatrix}$. In other words, given the type of the return distribution constant and the expected loss fixed at -10%, investors having a risk-aversion coefficient of *a*=4 are indifferent to the choice between a portfolio providing an expected return of -4% or 49%. This phenomenon might be surprising if variance is not taken into consideration (as in most of the experimental studies) and it underlines the importance of measuring risk-preference.

### 3.4 Robustness test and the time component

Furthermore, we provide a robustness test by running a linear regression model for annual, monthly and daily parameters as well. Here, we run regressions at daily, monthly and yearly return horizon for daily overlapping periods. The applied methodology consists of univariate OLS regressions, in which the predictor variables are volatility, 5% Value-at-Risk, 5% Conditional Value-at-Risk, CAPM Beta and EDR respectively. As noted before, the filtered sample for individual shares consists of 340 assets (i.e. 340 expected returns, and measures of risk accordingly), whereas for portfolios we use 10,000 randomized, long-only portfolios. The purpose of these regressions is to test whether EDR is important enough for investors to bear a risk premium, and whether it performs better against its competitors from standard univariate asset pricing models. In Table 2 we present the results of the simple linear model's estimating expected return.

**Please insert Table 2 here**

In contrast to the estimated positive relationship between return and risk of the volatility-based model in any terms, changing the risk measure to *EDR* reveals an interesting pattern: expected return and risk are only positively correlated in the short run. Regression on annual and monthly statistics indicate a positive relationship between *EDR* and *E(r)* both for individual assets and portfolios. This means that as the expectation of the negative outcomes increases (the risk decreases), the expected return increases as well, which is against the findings of standard asset pricing theories. Furthermore, analyzing the p-values reveals that pricing regressions indicate that investors indeed seem to focus more on the expected loss instead of the simple risk measures as volatility and variance. Moreover, as the p-value is lower in *EDR* regressions than in *VaR* and *CVaR* models, it seems that one considers the losses on the whole domain instead of taking into account the tail risk only (e.g. at the 5% probability level analyzed). *EDR* surpasses in goodness-of-fit the widely used CAPM and its *Beta* measure as well in five out of the six tests, however, the *Beta* yields much worse, insignificant estimations for individual assets. Altogether, we conclude that *EDR* seems to be a more precise estimator of *E(r)* than its competitors.

In order to analyze the effect of the period length on the leveraged portfolio line, we run linear regressions for distinct intervals and test whether the coefficients between risk and return show a robust pattern. Figure 8 represents our results indicating a robust positive relationship between the coefficient of *EDR* and the length of the analyzed period. Again, this phenomenon implies that focusing on the very short term, investors seem to be risk-averse and at around 11-12 days they are insensitive to risk; for periods over 12 days a negative relationship between risk and return applies to them. Here, we highlight again that this risk-seeking behavior exists in the sense that portfolios with different expected return may provide the same utility for given expected downside risk;

however, it does not reject the Expected Utility Theory or require negative risk-aversion coefficients.

**Please insert Figure 8 here**

## 4. Concluding remarks

Numerous studies found evidence of investors' behavior contradicting the well-known positive risk-reward relationship. However, asset pricing theories based on the expected utility theory have not yet given any explanation of a negative relationship under standard circumstances. Experimental evidences show that the risk-aversion coefficient is positive for a dominant portion of investors (and hence they are risk-averse), although, in some cases they systematically behave in a seemingly risk-seeking way.

In this paper we argue that this deviation from the expected behavior might be due to the false definition of risk-perception instead of a flaw in the definition of the perceived utility (as previous studies suggest). Therefore, in our model we propose the Expected Downside Risk as an alternative risk measure that describes and better fits investors' sensitivity to risk as measured by pricing regressions. Being a simple measure of the average loss relative to the expected return *EDR* seems to yield a more perceived value of risk than the standard deviation from the historical mean, especially in the presence of a highly discrete number of choices (e.g. experimental tests). This latter statement is confirmed by our regression results as well, where expected return is more significantly driven by *EDR* than volatility in all three period lengths for both individual assets and portfolios.

Having confirmed that *EDR* measures risk-perception better than volatility; asset pricing models based on expected utility theory can be modified by replacing the latter with the former risk measure. The solution of the optimization problem reveals that both a positive and negative relationship between expected return and *EDR* may exist under standard circumstances.

Finally, our model is supported by regression results indicating a negative relationship between risk and expected return for periods over 12 days and the "usual" positive relationship for less than 12 days.

For further research one may include the test of Expected Downside Risk in experimental settings, the detailed analysis of the effect of time on the risk-return relationship measured by *EDR* or by examining portfolio optimization where both negative and positive relationship can be found depending on the exogenous parameters, the latter of which could yield an alternative method for measuring risk-aversion.

## Table 1: Utilities at different risk-aversion for fixed *EDR*=0.15

|  | a=0.5 | a=4 | a=10 |
|---|---|---|---|
| E(U\|(E(r)=0.27) | 0.2641 | 0.2224 | 0.1510 |
| E(U\|E(r)=0.24) | 0.2367 | 0.2132 | 0.1731 |

Notes: Table 1 stands for the expected utilities of the pricing equation $U = E(r(x)) - \frac{0.5}{v^2}a[E(r(x)) - EDR(x)]^2$ given *EDR*=0.15, *E(r)*=0.27 and *E(r)*=0.24 for a=0.5, a=4 and a=10 risk-aversion coefficients.

## Table 2: Linear estimations of the expected return

|  |  | Individual shares | | | | | Portfolios | | | | |
|---|---|---|---|---|---|---|---|---|---|---|---|
|  |  | σ | VaR(5%) | CVaR(5%) | Beta | EDR | σ | VaR(5%) | CVaR(5%) | Beta | EDR |
| Yearly | Coeff | 2.7E-02 | 3.5E-02 | 2.0E-02 | 3.3E-04 | 1.6E-01 | 4.7E-02 | 6.5E-02 | 2.3E-02 | 7.9E-03 | 2.0E-01 |
|  | p-value | 1.4E-01 | 1.2E-01 | 1.7E-03 | 9.3E-01 | 8.3E-15 | 2.5E-35 | 6.7E-28 | 2.2E-55 | 8.2E-25 | 0.0E+00 |
| Monthly | Coeff | 7.2E-03 | 2.7E-02 | 3.2E-03 | -2.5E-04 | 1.2E-02 | 7.3E-03 | 2.3E-02 | 3.9E-03 | 1.1E-03 | 2.1E-02 |
|  | p-value | 2.9E-01 | 3.6E-01 | 2.3E-01 | 6.2E-01 | 2.2E-01 | 9.1E-08 | 1.3E-03 | 1.6E-12 | 4.6E-26 | 4.8E-28 |
| Daily | Coeff | 3.0E-03 | 6.2E-02 | -1.1E-03 | 4.3E-06 | -6.1E-03 | 2.1E-03 | 4.1E-02 | -4.2E-04 | 8.6E-05 | -4.6E-03 |
|  | p-value | 4.9E-02 | 4.0E-02 | 1.2E-01 | 9.0E-01 | 1.5E-02 | 1.0E-11 | 3.4E-08 | 2.7E-03 | 3.3E-35 | 3.0E-20 |

Notes: Table 2 represents the OLS estimations of the $\widehat{E(r)} = \hat{\alpha} + \hat{\beta} \cdot y$ where y stands for volatility (σ), 5% Value-at-Risk (VaR 5%), 5% Conditional Value-at-risk (CvaR 5%), CAPM Beta and EDR measures. The $\hat{\beta}$ coefficients and their significance levels (p-values) are shown for individual shares and randomly generated portfolios using yearly, monthly and daily returns.

# Figure 1: Iso-utility functions in EDR-E(r) system given a=4

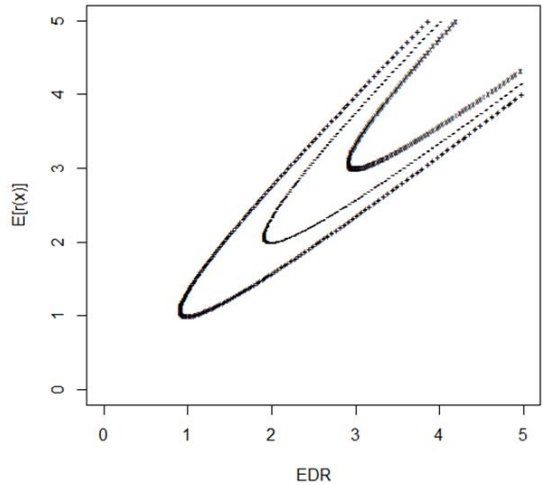

**Notes: Figure 1 represents iso-utility functions in the *EDR-E(r)* system given a=4 risk-aversion coefficient. The points signed by +, - and x represent utility levels of U=1, 2 and 3 respectively.**

# Figure 2: Limits of iso-utility functions in the EDR-E(r) system

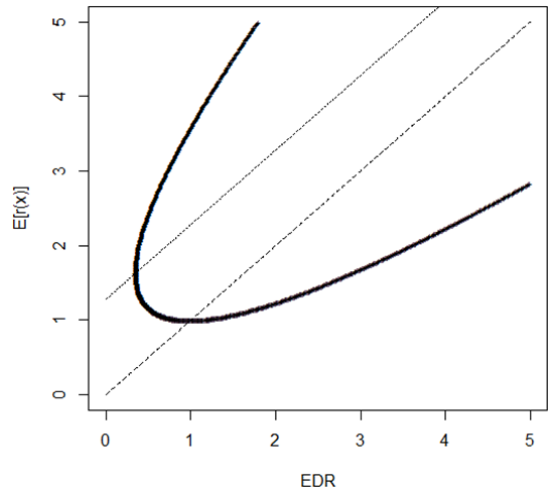

**Notes: Figure 2 represents the indifference curve (solid line), risk-free portfolios (dashed line), and the $EDR(x) = E(r(x)) - \frac{v^2}{a}$ constraint (dotted line) respectively, given a=0.5 risk-aversion coefficient and v=0.8 volatility coefficient.**

### Figure 3 here: Leveraged portfolio optimization

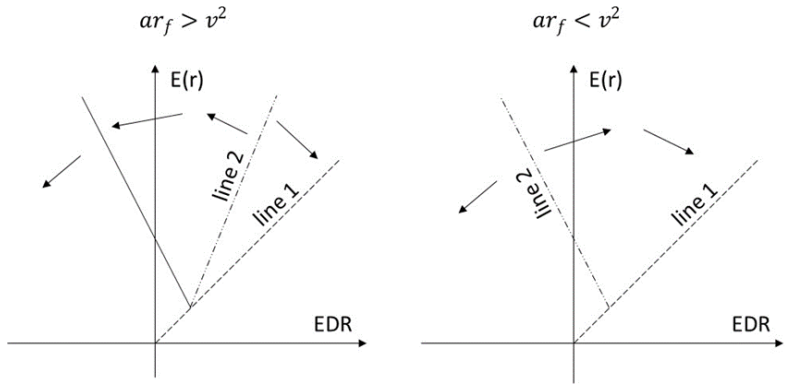

**Notes: Figure 3 represents the leveraged portfolio optimization where line 1 and line 2 stand for the $EDR \leq E(r)$ constraint and the $k = \frac{ar_f}{ar_f - v^2}$ condition (i.e. the slope where utility is minimal) respectively. Investors try to reach the leveraged portfolio line the furthest from line 2.**

### Figure 4: Unleveraged portfolios in EDR-E(r) system

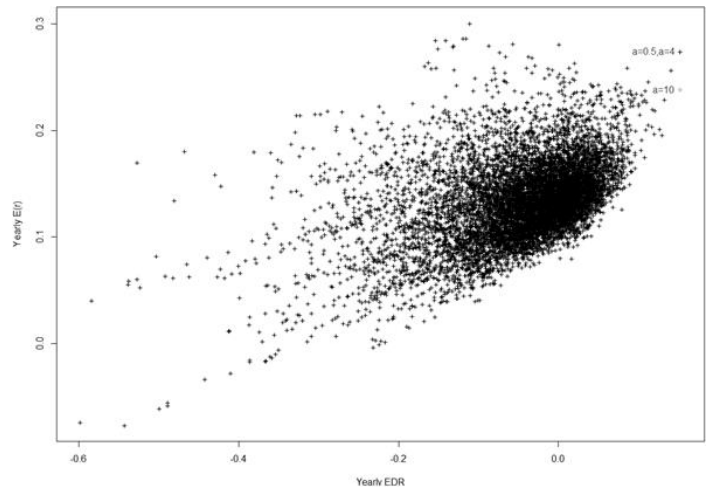

**Notes: Figure 4 represents the *EDR-E(r)* pairs of 10,000 randomly simulated portfolios of the 340 S&P500 members existent both at the beginning and at the end of the analyzed period. The gray points stand for the optimal portfolio (i.e. providing the highest expected utility) given a=0.5 or a=4 and a=10 risk-aversion coefficients. The parameters are calculated using non-overlapping yearly returns.**

## Figure 5: Unleveraged portfolios in daily analysis

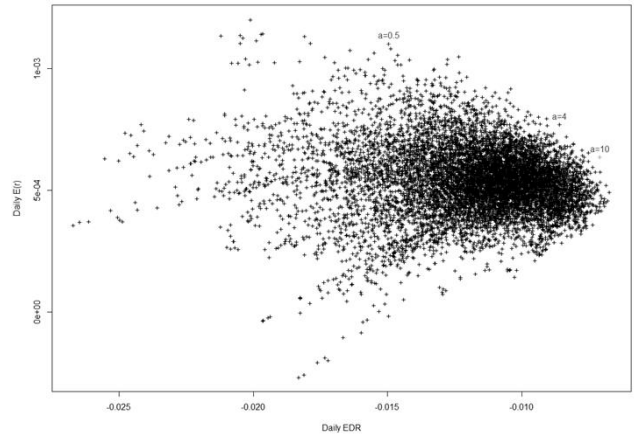

**Notes: Figure 4 represents the *EDR-E(r)* pairs of 10,000 randomly simulated portfolios of the 340 S&P500 members existent both at the beginning and at the end of the analyzed period. The gray points stand for the optimal portfolio (i.e. providing the highest expected utility) given a=0.5, a=4 and a=4 risk-aversion coefficients respectively. The parameters are calculated using non-overlapping daily returns.**

## Figure 6: Leveraged portfolios in EDR-E(r) system

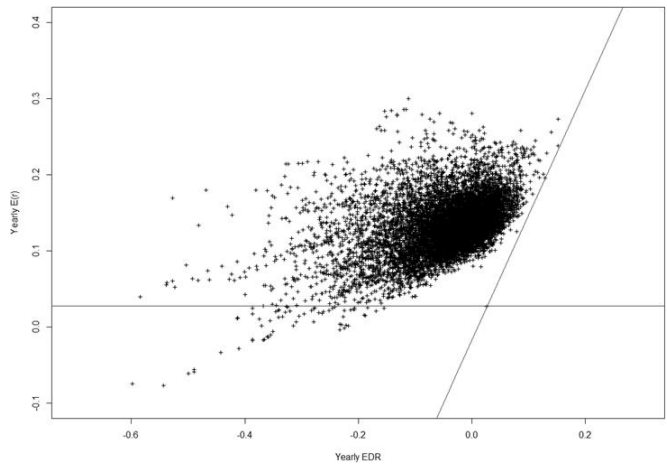

**Notes: Figure 6 shows the leveraged portfolio optimization conditional on yearly parameters, where the "+" sign in the intersection of the lines stands for the risk-free asset, while the horizontal and $\frac{dE(r(x))}{dEDR(x)} = 1.64$ sloped lines represent the optimal leveraged portfolio line given $a \geq 28.55$ and $a < 28.55$ respectively.**

**Figure 7: Leveraged portfolios in daily analysis**

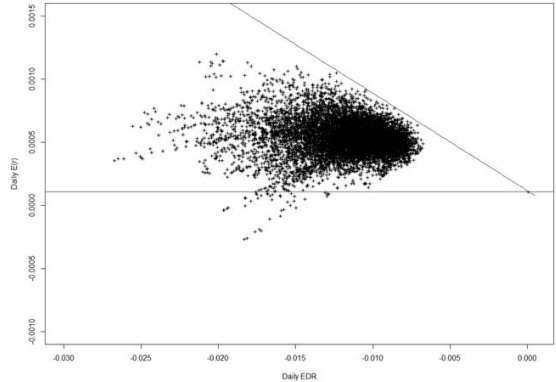

**Notes: Figure 7** shows the leveraged portfolio optimization conditional on daily parameters, where the "+" sign in the intersection of the lines stands for the risk-free asset while the horizontal and $\frac{dE(r(x))}{dEDR(x)} = -0.08$ sloped lines represent the optimal leveraged portfolio line given $a \geq 155.69$ and $a < 155.69$ respectively.

**Figure 8: Relationship between time, risk and return**

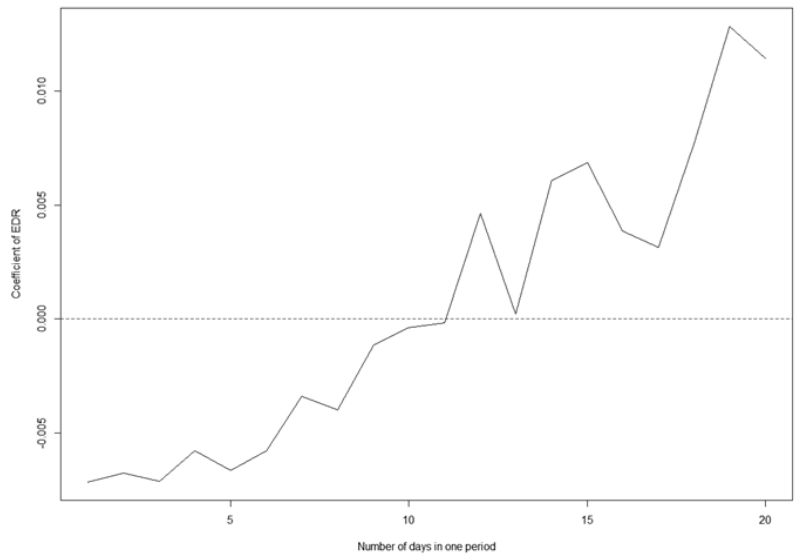

**Notes: Figure 8** represents horizon-dependent relationship between the β coefficient of EDR in the $\widehat{E(r)} = \hat{\alpha} + \hat{\beta} \cdot EDR$. This phenomenon implies that focusing on the very short term, investors seem to be risk-averse, and therefore, their coefficient is positive. However, at around 11-12 days they become insensitive to risk and for investment periods over 12 days a negative relationship between risk and return applies to them. The dashed line stands for the zero value.